\documentclass[twocolumn,showpacs,preprintnumbers,amsmath,amssymb]{revtex4}
\usepackage{color}
\usepackage{bm, graphicx, amsmath}
\usepackage{hyperref}

\begin{document}

\title{Discrete-outcome sensor networks: Multiple detection events and grouping detectors}
\author{Nada Ali$^{1,2}$ and Mark Hillery$^{1,2}$}
\affiliation{$^{1}$Department of Physics, Hunter College of the City University of New York, 695 Park Avenue, New York, NY 10065 USA \\ $^{2}$Physics Program, Graduate Center of the City University of New York, 365 Fifth Avenue, New York, NY 10016}

\begin{abstract}
Quantum sensor networks have often been studied in order to determine how accurately they can determine a parameter, such as the strength of a magnetic field, at one of the detectors.  A more coarse-grained approach is to try to simply determine whether a detector has interacted with a signal or not, and which detector it was.  Such discrete-outcome quantum sensor networks, discrete in the sense that we are seeking answers to yes-no questions, are what we study here. One issue is what is a good initial state for the network, and, in particular, should it be entangled or not. Earlier we looked at the case when only one detector interacted, and here we extend that study in two ways.  First, we allow more that one detector to interact, and second, we examine the effect of grouping the detectors.  When the detectors are grouped we are only interested in which group contained interacting detectors and not in which individual detectors within a group interacted.  We find that in the case of grouping detectors, entangled initial states can be helpful.
\end{abstract}

\maketitle

\section{Introduction}

Can quantum mechanics, in particular entanglement, be used to improve the behavior of sensor networks?  Quantum effects, in particular nonclassical states of light, can be used to improve the performance of individual detectors, with the most spectacular example being the use of squeezed states to improve the sensitivity of the LIGO gravitational wave detector.  In a network of detectors there are additional possibilities resulting from entanglement and the possibility of global measurements.  So far it is not clear when, and if, these can lead to improved sensitivity. 
		
For the most part, previous work on quantum sensor networks has studied the following problem.  The detectors, which are quantum systems, have interacted with the environment, and, as a result, parameters due to this interaction are encoded in their state.  In almost all cases, these parameters have been taken to be continuous variables.  For example, these parameters could be the strength of a magnetic field at different locations.  The object is to estimate these parameters or some function of them.  The detectors can be of various types, qubits \cite{gorshkov,qian}, continuous variable systems \cite{shapiro}, or general quantum systems \cite{proctor}.  It has been found that for finite-dimensional systems, entanglement of the quantum systems does not provide an advantage in estimating the individual parameters, but does provide an advantage in estimating a function of them \cite{gorshkov,qian,proctor,Rubio}.  It has been shown that entangled states in optical networks can provide an advantage for distributed sensing \cite{shapiro,Zhang}.  Further studies have investigated whether linear optical networks with unentangled inputs can give a quantum advantage in distributed metrology \cite{ge}, and whether continuous-variable error correction can be useful in protecting a network of continuous-variable sensors from the effects of noise \cite{preskill}.
	
In previous papers we looked at a different problem \cite{hillery,zhan}.  Suppose that instead of determining a parameter, one is interested in whether a detector has detected something or which detector has detected something.  This kind of problem is described by discrete rather than continuous variables, and is a problem in channel discrimination \cite{kitaev,acin,DAriano,sacchi1,sacchi2,wang,pirandola}.  Each detector in a network can receive an input or no input.  Suppose the unitary operator $U$ describes the interaction between an input and a detector, and that only one detector in the network has received an input, but we do not know which one.  This is the problem we studied in \cite{hillery}.  The operator $U$ could describe, for example, the rotation of a spin caused by a magnetic field, or a phase shift induced in a state of light by a transparent object.  The different output states of the detectors will be produced by starting with the initial state of the detectors and applying to it the operator consisting of $U$ acting on one of the detectors and the identity acting on the rest.  We then want to measure the output state in order to determine which detector received an input.  This means that we have to optimize over both the initial state of the detectors, and the final measurement.  What we found was that for a small number of detectors starting from an initially entangled state helps, but the advantage decreases with the number of detectors.  A related problem of picking out a target quantum channel from a background of identical channels has been analyzed by Zhuang and Pirandola, and useful bounds on channel discrimination have been derived \cite{Zhuang1,Zhuang2,Pereira}.
	
Here we would like to remove the restriction that only one of the detectors interacted with the environment and allow an arbitrary number of detectors to interact.  We will begin by analyzing a two-detector network.  We will look at two measurement schemes for the final state of the detectors, minimum  error and unambiguous \cite{helstrom,review}.  In minimum-error detection there is a probability of making a mistake, but this probability is minimized.  In unambiguous discrimination, one never makes a mistake, but the measurement can sometimes fail.  Which strategy to use depends on the relative cost assigned to making a mistake versus not getting an answer.

Next we will look at the situation where detectors are grouped, and all we are interested in is in which group detectors fired and not in which individual detectors fired. For example, if the detectors are in two groups, group 1 and group 2, all we wish to know is whether some detectors in group 1 fired or some detectors in group 2 fired.  This type of situation could arise if the group 1 detectors are localized in one region and the group 2 detectors are in another, and we are just trying to detect in which region an event occurs.  Here there are two issues.  The first is whether an entangled initial state helps.  In this case we will find that it does.  A second question is whether collective measurements, in which several detectors are measured together instead of individually, helps.  We provide and example for the case of unambiguous discrimination in which it does. Grouping detectors was studied in \cite{gupta}, where it was employed as part of a two-part method to localize a single detection event; first find the group in which the detection occurred, and then find the individual detector within the group at which the event occurred.

In the case of grouped detectors, we will examine several different scenarios, but all will take as their object of study two groups of two detectors each.  We will first use minimum-error discrimination.  In the case when only one detector in each group can fire, we show that an entangled initial state can be used to reduce this problem to one of only two detectors.  Next, we will allow more than one detector in each group to fire, and compare the performance of an entangled initial state and a product initial state.  We will then look at this system using unambiguous discrimination with an initial product state, and show that a collective measurement performs better than measuring each detector individually.

 \section{Two detectors}
 In order to begin we will start with a very simple system, a system consisting of two detectors, and any combination of them can interact with the environment, neither, both, or just one of the two.  The interaction is described by a unitary operator $U$, with eigenstates $|u_{\pm}\rangle$, where $U|u_{\pm}\rangle = e^{\pm i\theta} |u_{\pm}\rangle$.  The parameter $\theta$ is a product of the interaction strength between the detector and the environment and the interaction time.  If one is trying to detect weak signals, $\theta$ will be small.  If $|\psi_{in}\rangle$ is the initial state of the detectors, the state afterwards will be one of the states, $|\psi_{in}\rangle$, $(U\otimes I)|\psi_{in}\rangle$, $(I\otimes U)|\psi_{in}\rangle$, or $(U\otimes U)|\psi_{in}\rangle$.  
 
 We now need to choose an initial state of the detectors, and we will give a plausibility argument, which is not a proof, for making a particular choice.   First let us note that what we have here is a channel discrimination problem.  The four channels are $I\otimes I$, $I\otimes U$, $U\otimes I$, and $U\otimes U$.  The initial state $|\Psi_{in}\rangle$ should be chosen so that the resulting output states are as distinguishable as possible.  That means we want to minimize the overlaps between different output states.  If their overlaps were zero, the states would be perfectly distinguishable, and we could determine with certainty which state we had.  However, in general this cannot be arranged, and we will have to accept some probability of error or some probability of failure depending on whether we adopt the minimum-error or unambiguous strategy.  Both the probability of error and the probability of failure depend on the overlaps of the set of states we are considering, and, in general, the smaller the overlaps the smaller the probabilities of error or failure.  Consequently, looking at the overlaps of the states corresponding to different combinations of detectors firing will give us an idea of what a good initial state is.  
 
 Let us first note that linear combinations of the states $\{ |u_{+}u_{+}\rangle ,\, |u_{-}u_{-}\rangle \}$ cannot distinguish between $I\otimes U$ and $U\otimes I$, while linear combinations of $\{ |u_{+}u_{-}\rangle ,\, |u_{-}u_{+}\rangle\}$ cannot distinguish between $I\otimes I$ and $U\otimes U$.  Similarly, linear combinations of of the states $\{ |u_{+}u_{+}\rangle ,\, |u_{+}u_{-}\rangle\}$ cannot distinguish between $I\otimes I$ and $U\otimes I$ (this is also true if the first state in each product state is $|u_{-}\rangle$ instead of $|u_{+}\rangle$), while linear combinations of $\{ |u_{+}u_{+}\rangle , \, |u_{-}u_{+}\rangle\}$ cannot distinguish between $I\otimes I$ and $I\otimes U$.  This suggests that $|\psi_{in}\rangle$ should be a linear combination of all four of the states $|u_{\pm}u_{\pm}\rangle$.  We should then find which choices of the coefficients in the linear combination minimizes the overlaps between different output states.  
 
Starting from the initial state
\begin{equation}
|\psi_{in}\rangle = \sum_{j=\pm} \sum_{k=\pm} c_{jk} |u_{j}u_{k}\rangle ,
\end{equation}
and setting 
\begin{eqnarray}
|\psi_{out0}\rangle = |\psi_{in} \rangle & \hspace{3mm} & |\psi_{out1}\rangle = (I\otimes U)|\psi_{in}\rangle \nonumber \\
|\psi_{out2}\rangle = (U\otimes I) |\psi_{in}\rangle & \hspace{3mm} & |\psi_{out3}\rangle = (U\otimes U)|\psi_{in}\rangle ,
\end{eqnarray}
we want to examine the overlaps between these four output states.  We find that $|\langle \psi_{out1}|\psi_{in}\rangle |^{2}$, $|\langle \psi_{out2}|\psi_{in}\rangle |^{2}$, $|\langle \psi_{out1}|\psi_{out3}\rangle |^{2}$, and $|\langle \psi_{out2}|\psi_{out3}\rangle |^{2}$ are all of the form $|ae^{i\theta} + be^{-i\theta}|^{2}$, where $0\leq a \leq 1$, $0\leq b \leq 1$, and $a+b=1$.  It is relatively straightforward to show that this expression is minimized when $a=b=1/2$.  This implies that all of these overlaps are minimized when $|c_{jk}|^{2} = 1/4$, where $j=\pm$ and $k=\pm$.  The remaining overlaps, $|\langle \psi_{out2}|\psi_{out1}\rangle |^{2}$ and $|\langle \psi_{out3}|\psi_{in}\rangle |^{2}$ are not minimized for the same values of the $c_{jk}$, but their average value is minimized when $|c_{jk}|^{2} = 1/4$.  Consequently, the suggestion is that the choice $c_{jk}=1/2$ will be a good one for producing four states that are close to optimally distinguishable.  This implies that the choice of $|\psi_{in}\rangle$ as a product state is a good one.  Note that what we have is not a proof that this is the best choice but a strong plausibility argument.  In particular, defining
 \begin{eqnarray}
 |w_{0}\rangle & = & \frac{1}{\sqrt{2}} ( |u_{+}\rangle + |u_{-}\rangle ) \nonumber   \\
 |w_{1}\rangle & = & \frac{1}{\sqrt{2}} ( e^{i\theta} |u_{+}\rangle + e^{-i\theta} |u_{-}\rangle ) ,
 \end{eqnarray}
 we choose $|\Psi_{in}\rangle = |w_{0}\rangle \otimes |w_{0}\rangle$, and the remaining states we must discriminate are $|w_{1}\rangle \otimes |w_{0}\rangle$, $|w_{0}\rangle \otimes |w_{1}\rangle$, and $|w_{1}\rangle \otimes |w_{1}\rangle$, which we shall denote by $|\phi_{jk}\rangle = |w_{j}\rangle \otimes |w_{k}\rangle$, where $j,k=0,1$.
 
\subsection{Minimum-error discrimination}
 It is impossible to discriminate nonorthogonal quantum states perfectly.  One is then faced with finding an imperfect strategy for doing so.  One is the minimum-error strategy in which there is a non-zero probability of making a mistake, but this probability is minimized.  A second strategy, unambiguous discrimination will be discussed in the next section.  It is only possible to find an optimal minimum-error strategy for the case of two states \cite{helstrom} or for special sets of more than two states.  
 
 The procedure for discriminating two, in general, mixed, states, $\rho_{0}$, which occurs with probability $p_{0}$, and $\rho_{1}$, which appears with probability $p_{1}$ is due to Helstrom \cite{helstrom}.  The optimal POVM can be constructed from the eigenstates of the operator $\Lambda = p_{1}\rho_{1} - p_{0}\rho_{0}$, and the probability of successfully discriminating the states is given by
 \begin{equation}
 P_{s} = \frac{1}{2}+ \frac{1}{2} \| \Lambda \|_{1}
 \end{equation}
 where the norm is the trace norm.  The POVM operators in the case that $\rho_{0}$ and $\rho_{1}$ are pure qubit states are the orthogonal projections onto the eigenstates of $\Lambda$.  There is no known solution for the general case of more than two states.  In the case that the set of states has a particular symmetry, optimal solutions can be found, but in general only approximate analytical solutions are available.
 
We will assume that $|w_{0}\rangle$ occurs with a probability of $p_{0}$ and $|w_{1}\rangle$ occurs with a probability of $p_{1}$.  In order to find the POVM that discriminates these states with a minimum error, as noted in the previous paragraph, we construct the operator
\begin{equation}
\Lambda = p_{1}|w_{1}\rangle\langle w_{1}| - p_{0} |w_{0}\rangle\langle w_{0}| 
\end{equation}
and find its eigenvectors and eigenvalues \cite{helstrom}.  We find that the eigenvalues are
\begin{equation}
\lambda = \frac{1}{2} \left\{ p_{1}-p_{0} \pm [p_{1}^{2} + p_{0}^{2} - 2p_{0}p_{1}\cos\theta ]^{1/2} \right\}.
\end{equation}
The eigenvectors are $(1/\sqrt{2})(|u_{+}\rangle \pm e^{i\alpha} |u_{-}\rangle$ where
\begin{equation}
e^{i\alpha}=\frac{p_{1}e^{-2i\theta} - p_{0}}{|p_{1}e^{2i\theta} - p_{0}|} .
\end{equation}
Note that this implies that $\alpha$ will be negative if $0 < \theta \leq \pi /4$.  The POVM operators are 
\begin{eqnarray}
\Pi_{0} & = & \frac{1}{2} (|u_{+}\rangle - e^{i\alpha} |u_{-}\rangle  )(\langle u_{+}| - e^{-i\alpha} \langle u_{-}|) \nonumber \\
\Pi_{1} & = & \frac{1}{2} (|u_{+}\rangle + e^{i\alpha} |u_{-}\rangle  )(\langle u_{+}| + e^{-i\alpha} \langle u_{-}|) .  \nonumber 
\end{eqnarray}
The conditional detection probabilities, $p(w_{j}|w_{k})$, where $j,k=0,1$, which are the probabilities that if the state is $|w_{k}\rangle$ then $|w_{j}\rangle$ is detected are
\begin{eqnarray}
p(w_{0}|w_{0}) & = & \langle w_{0}|\Pi_{0}|w_{0}\rangle = \frac{1}{2}(1- \cos\alpha ) \nonumber \\
p(w_{1}|w_{0}) & = & \langle w_{0}|\Pi_{1}|w_{0}\rangle = \frac{1}{2}(1+ \cos\alpha ) \nonumber \\
p(w_{0}|w_{1}) & = & \langle w_{1}|\Pi_{0}|w_{1}\rangle = \frac{1}{2}(1- \cos (\alpha + 2\theta ) )\nonumber \\
p(w_{1}|w_{1}) & = & \langle w_{1}|\Pi_{1}|w_{1}\rangle = \frac{1}{2}(1+ \cos (\alpha + 2\theta ) ) .
\end{eqnarray}
The probability of successfully discriminating the states is
\begin{equation}
P_{s}^{(1)} = p(w_{0}|w_{0}) p_{0} + p(w_{1}|w_{1}) p_{1} = \frac{1}{2} ( 1 + |p_{1}e^{2i\theta} - p_{0}|) .
\end{equation}

Proceeding now to the case of two two-qubit detectors, the POVM elements are $\Pi_{jk}= \Pi_{j}\otimes \Pi_{k}$, where $j,k=0,1$, and, assuming the two-qubit state $|w_{j}\rangle \otimes |w_{k}\rangle$ appears with probability $p_{j}p_{k}$, the success probability is
\begin{equation}
P_{s}^{(2)} = \sum_{j,k=0}^{1} p(w_{j}|w_{j}) p(w_{k}|w_{k}) p_{j} p_{k} = \left( P_{s}^{(1)}\right)^{2} .
\end{equation}
This result clearly generalizes to $N$ detectors if we assume an initial product state and the probabilities that the detectors fire are independent.  In the case that the probabilities that the states appear are not independent, that is that the probability that the state $|w_{j}\rangle \otimes |w_{k}\rangle$ appears is not $p_{j}p_{k}$, then, as was seen in our previous work the situation is different \cite{hillery}.  In that case only the states $|w_{0}\rangle \otimes |w_{1}\rangle$ and $|w_{1}\rangle\otimes |w_{0}\rangle$ could appear, each with a probability of $1/2$, and then the optimal initial state was entangled and the optimal measurements were global measurements that were projections onto entangled states.  This suggests that entanglement can provide an advantage when the sets of detectors that will fire possess correlations.

 \subsection{Unambiguous discrimination}
We will now look at the unambiguous discrimination of the states $|w_{j}\rangle\otimes |w_{k}\rangle$, where $j,k=0,1$.  In this measurement scheme we will never make a mistake, but the measurement may fail.  In the case of pure states, in order for unambiguous discrimination to be possible, the states must be linearly independent.  In the case of mixed states, the supports of the density operators must not be identical.  In our case the POVM elements will be of the form $\Pi_{jk}=c |\phi_{jk}^{\perp}\rangle\langle \phi_{jk}^{\perp}|$, where the vector $|\phi_{jk}^{\perp}\rangle$ is the unit vector that is orthogonal to $|w_{m}\rangle |w_{n}\rangle$ for $m\neq j$ or $n\neq k$, and the constant $c$ is chosen to be as large as possible so that the POVM element corresponding the the failure of the measurement, 
\begin{equation}
\Pi_{f}= I -c\sum_{j,k=0}^{1} |\phi^{\perp}_{jk}\rangle\langle\phi^{\perp}_{jk}| , 
\end{equation}
is positive.  Note that the POVM element $\Pi_{jk}$ corresponds to detecting the state $|w_{j}\rangle\otimes |w_{k}\rangle$, and the fact that $\Pi_{jk}(|w_{m}\rangle\otimes |w_{n}\rangle ) = 0$ for $m\neq j$ or $n\neq k$ implies that we will never make an error.

Define the vectors $|v_{0}\rangle = (1/\sqrt{2})(|u_{+}\rangle - |u_{-}\rangle )$ and $|v_{1}\rangle = (1/\sqrt{2})(e^{i\theta}|u_{+}\rangle - e^{-i\theta}|u_{-}\rangle )$.  Note that $\langle v_{j}|w_{j}\rangle =0$ for $j=0,1$.  We can now take
\begin{equation}
|\phi_{kl}^{\perp}\rangle = |v_{\bar{k}}\rangle |v_{\bar{l}}\rangle ,
\end{equation}
where $\bar{k} = k+1$ mod 2, and similarly for $\bar{l}$.  The failure operator can now be expressed as
\begin{equation}
\Pi_{f} = I - c \left(\sum_{j=0}^{1} |v_{j}\rangle\langle v_{j}|\right) \otimes \left(\sum_{k=0}^{1} |v_{k}\rangle\langle v_{k}|\right) .
\end{equation}
Defining $T = \sum_{j=0}^{1} |v_{j}\rangle\langle v_{j}| $, we see that if $\lambda_{max}$ is the largest eigenvalue of $T$, then $\Pi_{f}$ will be positive if $c\lambda_{max}^{2} \leq 1$, so we will set $c=1/\lambda_{max}^{2}$.  Now in matrix form we have that
\begin{equation}
T= \frac{1}{2} \left(\begin{array}{cc} 2 & -(e^{2i\theta} + 1) \\ -(e^{-2i\theta} + 1) & 2 \end{array}\right) ,
\end{equation}
so that $\lambda_{max} = 1 + (1/\sqrt{2})(1+\cos 2\theta )^{1/2} = 1+\cos\theta$ for $0\leq \theta \leq \pi /2$.  Assuming the states are equally likely, the failure probability is 
\begin{eqnarray}
P_{f} & = & 1 - \frac{c}{4} \sum_{k=0}^{1} \sum_{l=0}^{1} |\langle w_{k}|v_{\bar{k}}\rangle |^{2} |\langle w_{l}|v_{\bar{l}}\rangle |^{2} \nonumber \\
&=& 1-\frac{1}{4} \left( \frac{1}{\lambda_{max}} \sum_{k=0}^{1} |\langle w_{k}|v_{\bar{k}}\rangle |^{2} \right)^{2} .
\end{eqnarray}
We also have that $\langle v_{1}|w_{0}\rangle = \langle v_{0}|w_{1}\rangle = i\sin\theta$, so 
\begin{equation}
P_{f}=1- \frac{\sin^{4}\theta}{ [ 1+\cos\theta ]^{2}} .
\end{equation}
This is plotted in Fig. 1. As can be seen, the failure probability is close to one for small angles.  The reason for this is that the second term in the above equation is proportional to $\theta^{4}$ for small angles.  The success probability is $1-P_{f}$, and is just the probability that both measurements succeed.

This procedure can easily be generalized for more than two detectors if we start from a product state $|\psi_{in}\rangle = |w_{0}\rangle^{\otimes n}$.  We will then be faced with discriminating states of the form $|\phi_{x}\rangle = |w_{x_{n-1}}\rangle |w_{x_{n-2}}\rangle \ldots |w_{x_{0}}\rangle$, where $x=x_{n-1}x_{n-2} \ldots x_{0}$ is an $n$-digit binary number.  The POVM element corresponding to detecting this state is $\Pi_{x} = c |\phi^{\perp}_{x}\rangle \langle\phi^{\perp}_{x}|$, where $|\phi^{\perp}_{x}\rangle = |v_{\bar{x}_{n-1}}\rangle \ldots |v_{\bar{x}_{0}}\rangle$.  The constant $c$ is now given by $c=1/\lambda_{max}^{n}$, and the failure probability for $0\leq \theta \leq \pi /2$, if the states are equally probable, is
\begin{equation}
\label{unam-n}
P_{f} = 1- \frac{\sin^{2n}\theta}{ [1+\cos \theta ]^{n}} .
\end{equation}

\begin{figure}
\label{figure1}
\centering
\includegraphics[width=0.5\textwidth]{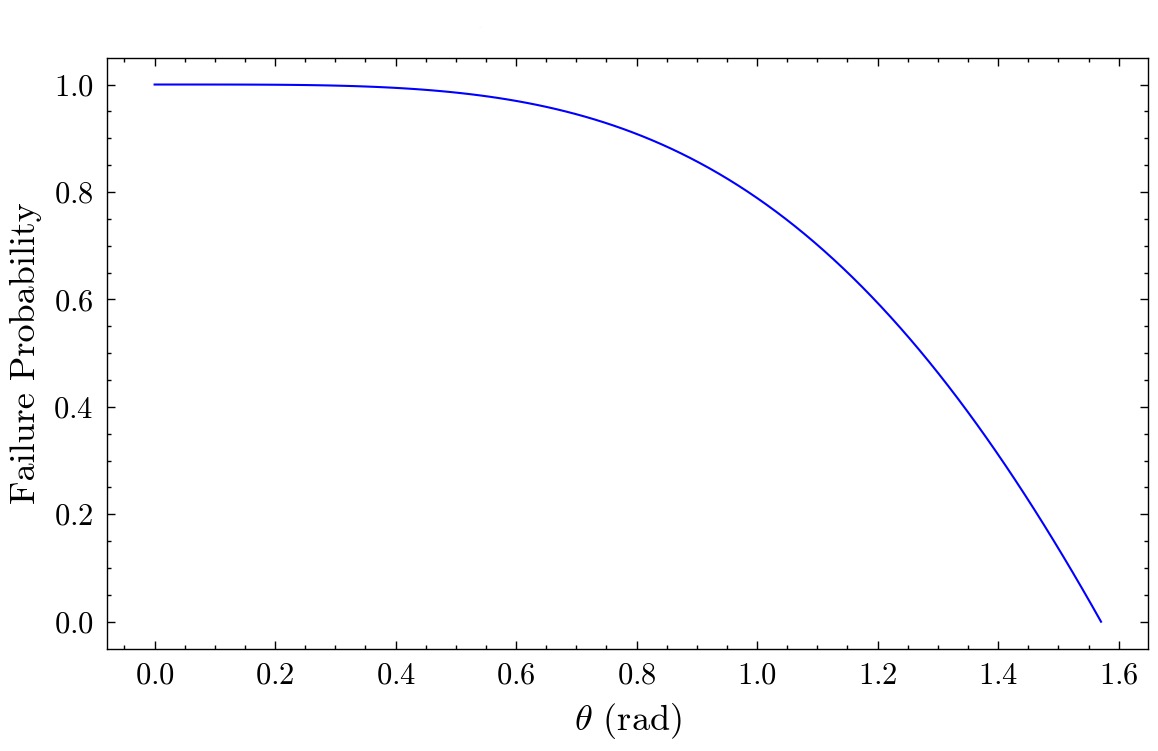}
\caption{Failure probability, $P_{f}$ as a function of $\theta$.  The failure probability remains close to one for a significant range of $\theta$, suggesting that the unambiguous measurement strategy is not an optimal one for this situation.}
\end{figure}

\section{Grouping detectors}
Suppose we are only interested in whether detectors in certain groups fire.  In particular, suppose we divide our detectors into two sets, and all we are interested in is whether some detectors in the first set fired, or whether some in the second set fired.  We will look at two cases.  In the first, only a single detector will fire, and in the second multiple detectors in a single group can fire.

\subsection{Single detector fires}

We will first consider the case in which only a single detector fires, and we will examine the case of four detectors split into two groups of two.  However, it is first useful to review the case of only two detectors when it is guaranteed that only one of them fires \cite{hillery}.  We start with the state
\begin{equation}
\label{2det-1f}
\frac{1}{\sqrt{2}} \left( |u_{+}\rangle_1 |u_{-}\rangle_2 + |u_{-}\rangle_1 |u_{+}\rangle_2 \right) .
\end{equation}
If detector 1 fires we apply $ U \otimes I $ to this state and get
\begin{equation}
\frac{1}{\sqrt{2}} \left( e^{i\theta} |u_{+}\rangle_1 |u_{-}\rangle_2 + e^{-i\theta} |u_{-}\rangle_1 |u_{+}\rangle_2  \right) .
\end{equation}
If, instead, detector 2 fires, we apply $I \otimes U$ to the state and we get 
\begin{equation}
\frac{1}{\sqrt{2}} \left( e^{-i\theta} |u_{+}\rangle_1 |u_{-}\rangle_2 + e^{i\theta} |u_{-}\rangle_1 |u_{+}\rangle_2 \right).
\end{equation}
Our task, then, is to distinguish these states, and we will present the results for minimum-error discrimination (see \cite{hillery} for the case of unambiguous discrimination).  The optimal measurements are global ones, that is, both qubits are measured together, not individually, and the operators describing the measurements are proportional to projections onto entangled states.  The measurement operators are orthogonal projections and are given by
\begin{eqnarray} 
\label{2state-minerr}
\Pi_{1} & = & |v_{1}\rangle \langle v_{1}| \nonumber \\
\Pi_{2} & = & |v_{2}\rangle \langle v_{2}| .
\end{eqnarray}
where
\begin{eqnarray}
|v_{1}\rangle & = & \frac{1}{\sqrt{2}} (|u_{+}\rangle_{1} |u_{-}\rangle_{2} -i |u_{-}\rangle_{1} |u_{+}\rangle_{2} ) \nonumber \\
|v_{2}\rangle & = & \frac{1}{\sqrt{2}} (|u_{+}\rangle_{1} |u_{-}\rangle_{2} + i |u_{-}\rangle_{1} |u_{+}\rangle_{2} ) .
\end{eqnarray}
The success probability, that is, the probability of getting the right answer, is
\begin{equation}
P_{s}^{(min)} = \frac{1}{2}[ 1 + \sin (2\theta )] ,
\end{equation}
where we have assumed that the two states are equally likely. 

Now let us move on to two groups of two detectors.  The first group contains detectors 1 and 2, and the second group contains detectors 3 and 4.  We are only considering the case where one detector in one of the groups fires, and we only want to know in which group the firing detector lies.  We will start with the state
\begin{equation}
\label{4det-1f}
\frac{1}{\sqrt{2}} \left( |u_{+}\rangle_1 |u_{+}\rangle_2 |u_{-}\rangle_3 |u_{-}\rangle_4 + |u_{-}\rangle_1 |u_{-}\rangle_2 |u_{+}\rangle_3 |u_{+}\rangle_4 \right) .
\end{equation}
Note that this state was obtained from the one in Eq.\ (\ref{2det-1f}) by replacing $|u_{+}\rangle$ by $|u_{+}\rangle |u_{+}\rangle$ and $|u_{-}\rangle$ by $|u_{-}\rangle |u_{-}\rangle$.  If a detector in the first group fires, then we apply $U \otimes I \otimes I \otimes I $ or $I \otimes U \otimes I \otimes I$ to this state obtaining, in both cases,
\begin{equation}
\frac{1}{\sqrt{2}} \left( e^{i\theta} |u_{+}\rangle_1 |u_{+}\rangle_2 |u_{-}\rangle_3 |u_{-}\rangle_4 + e^{-i\theta} |u_{-}\rangle_1 |u_{-}\rangle_2 |u_{+}\rangle_3 |u_{+}\rangle_4 \right)
\end{equation}
If a detector in the second group fires we apply $I \otimes I \otimes U \otimes I$ or $I \otimes I \otimes I \otimes U$ to the initial state resulting in
\begin{equation}
\frac{1}{\sqrt{2}} \left( e^{-i\theta} |u_{+}\rangle_1 |u_{+}\rangle_2 |u_{-}\rangle_3 |u_{-}\rangle_4 + e^{i\theta} |u_{-}\rangle_1 |u_{-}\rangle_2 |u_{+}\rangle_3 |u_{+}\rangle_4 \right) .
\end{equation}
Our problem is now to discriminate between these two states, but that is essentially the same as discriminating between the states in the case of two detectors.  We can obtain the POVM elements for the four-detector case from those of the two detector case by making the replacement $|u_{\pm}\rangle \rightarrow |u_{\pm}\rangle |u_{\pm}\rangle$.  This procedure clearly generalizes to more detectors in each group, and to a larger number of groups.  For the case in which only a single detector fires, it reduces the problem of discriminating among groups of detectors to discriminating among single detectors. 

\subsection{Multiple detectors fire}

Let us now consider four detectors split into two groups as before, but now either one detector or both detectors in a group can fire.  We will assume for now that the case in which detectors in both sets fire does not occur.  In this section we will be making use of minimum error detection.  The case of unambiguous discrimination will be examined in the next section.  If our initial state is $|\psi_{in}\rangle$, we are distinguishing between the sets of states 
\begin{eqnarray} 
\label{set1}
\{ (U\otimes I \otimes I \otimes I)|\psi_{in}\rangle, & \hspace{1mm} & (I\otimes U \otimes I \otimes I)|\psi_{in}\rangle ,\nonumber \\
 (U\otimes U \otimes I \otimes I)|\psi_{in}\rangle \}  && ,
\end{eqnarray}
and
\begin{eqnarray}
\label{set2}
\{ (I\otimes I \otimes U \otimes I)|\psi_{in}\rangle , & \hspace{1mm} & (I\otimes I \otimes I \otimes U) |\psi_{in}\rangle ,\nonumber \\ 
(I\otimes I \otimes U \otimes U)|\psi_{in}\rangle \} && .
\end{eqnarray}
As we shall see, each of these sets will correspond to a density matrix, $\rho_{12}$ for the first set and $\rho_{34}$ for the second.  Our task will then be to discriminate these two density matrices.  This same general structure would hold with larger numbers of detectors in each group.  We would again have two density matrices, each describing all of the ways within a given group can fire.  We would then have to discriminate these two density matrices.

We want to compare two initial states, an entangled state
\begin{equation}
|\psi_{in}\rangle = \frac{1}{\sqrt{2}}( |u_{+},u_{+}\rangle_{12} |u_{-},u_{-}\rangle_{34} + |u_{-},u_{-}\rangle_{12} |u_{+},u_{+}\rangle_{34}) ,
\end{equation}
and a product state
\begin{equation}
|\psi_{in}\rangle = \frac{1}{4} \prod_{j=1}^{4} (|u_{+}\rangle_{j} + |u_{-}\rangle_{j}) = |w_{0}\rangle^{\otimes 4} .
\end{equation}

In this section we will only be examining the case of minimum-error discrimination.  In order for unambiguous discrimination of two mixed states to be possible, it is necessary that their supports not be the same.  The next example we consider will not satisfy this requirement, so that unambiguous discrimination is not a possibility. 

Let us first look at the entangled state.  Applying the operators to the initial state results in two sets of states, which we would like to discriminate,
\begin{eqnarray}
\left\{ \frac{1}{\sqrt{2}} ( e^{i\theta} |u_{+},u_{+}\rangle_{12} |u_{-},u_{-}\rangle_{34} \right.  \nonumber \\ + e^{-i\theta} |u_{-},u_{-}\rangle_{12} |u_{+},u_{+}\rangle_{34}), \nonumber \\
\frac{1}{\sqrt{2}} ( e^{2i\theta} |u_{+},u_{+}\rangle_{12} |u_{-},u_{-}\rangle_{34} \nonumber \\
\left. + e^{-2i\theta} |u_{-},u_{-}\rangle_{12} |u_{+},u_{+}\rangle_{34}) \right\} , \nonumber \\
\end{eqnarray}
and
\begin{eqnarray}
\left\{ \frac{1}{\sqrt{2}} ( e^{-i\theta} |u_{+},u_{+}\rangle_{12} |u_{-},u_{-}\rangle_{34} \right. \nonumber \\
+ e^{i\theta} |u_{-},u_{-}\rangle_{12} |u_{+},u_{+}\rangle_{34}),  \nonumber \\
\frac{1}{\sqrt{2}} ( e^{-2i\theta} |u_{+},u_{+}\rangle_{12} |u_{-},u_{-}\rangle_{34} \nonumber \\
\left. + e^{2i\theta} |u_{-},u_{-}\rangle_{12} |u_{+},u_{+}\rangle_{34}) \right\} . \nonumber \\
\end{eqnarray}
Let us denote the two vectors in the first set by $|v_{1}\rangle$ and $|v_{2}\rangle$, and those in the second set by $|v_{3}\rangle$ and $|v_{4}\rangle$.  If we assume that each of the states in Eq.\ (\ref{set1}) are equally likely, that means $|v_{1}\rangle$ is twice as likely as $|v_{2}\rangle$, because it can occur in two different ways.  Similarly, $|v_{3}\rangle$ is twice as likely as $|v_{4}\rangle$.  Therefore, we can find a measurement that distinguishes between the two sets by finding one that distinguishes between the two density matrices
\begin{eqnarray}
\rho_{12} & = & \frac{2}{3} |v_{1}\rangle\langle v_{1}| + \frac{1}{3} |v_{2}\rangle\langle v_{2}| \nonumber  \\
\rho_{34} & = & \frac{2}{3} |v_{3}\rangle\langle v_{3}| + \frac{1}{3} |v_{4}\rangle\langle v_{4}|  .
\end{eqnarray}
The success probability in distinguishing between these two density matrices is given by the Helstrom formula
\begin{equation}
P_{s}=\frac{1}{2} + \frac{1}{4} \| \rho_{12} - \rho_{34} \|_{1} .
\end{equation}

In the $\{ |u_{+},u_{+}\rangle_{12} |u_{-},u_{-}\rangle_{34} ,\, |u_{-},u_{-}\rangle_{12} |u_{+},u_{+}\rangle_{34} \}$ basis these density matrices are
\begin{equation}
\rho_{12} = \left( \begin{array}{cc} \frac{1}{2} & \frac{1}{3} e^{2i\theta} + \frac{1}{6} e^{4i\theta} \\ \frac{1}{3} e^{-2i\theta} + \frac{1}{6} e^{-4i\theta} & \frac{1}{2} \end{array} \right) ,
\end{equation}
and
\begin{equation} 
\rho_{34} = \left( \begin{array}{cc} \frac{1}{2} & \frac{1}{3} e^{-2i\theta} + \frac{1}{6} e^{-4i\theta} \\ \frac{1}{3} e^{2i\theta} + \frac{1}{6} e^{4i\theta} & \frac{1}{2} \end{array} \right) .
\end{equation}
The eigenvalues of $\rho_{12}-\rho_{34}$ are $\lambda = \pm [(2/3)\sin (2\theta )+ (1/3) \sin (4\theta )]$.  For the trace norm we have
\begin{equation}
 \| \rho_{12} - \rho_{34} \|_{1} = 2\left[ \frac{2}{3} \sin (2\theta ) + \frac{1}{3} \sin (4\theta ) \right] ,
 \end{equation}
 and the success probability is
 \begin{equation}
 P_{s} = \frac{1}{2} +  \frac{1}{3} \sin (2\theta ) + \frac{1}{6} \sin (4\theta ) .
 \end{equation}
 
 Now let's look at the case when $|\psi_{in}\rangle$ is a product state.  Define the vectors
 \begin{eqnarray}
 |\mu_{1}\rangle = |w_{1}\rangle \otimes |w_{0}\rangle^{\otimes 3} &\hspace{4mm} &  |\mu_{2}\rangle = |w_{0}\rangle |w_{1}\rangle |w_{0}\rangle^{\otimes 2} \nonumber \\
 |\mu_{3}\rangle = |w_{1}\rangle^{\otimes 2} |w_{0}\rangle^{\otimes 2} & \hspace{4mm} & |\mu_{4}\rangle = |w_{0}\rangle^{\otimes 2} |w_{1}\rangle |w_{0}\rangle \nonumber \\
 |\mu_{5}\rangle =  |w_{0}\rangle^{\otimes 2} |w_{0}\rangle |w_{1}\rangle & \hspace{4mm} &  |\mu_{6}\rangle = |w_{0}\rangle^{\otimes 2} |w_{1}\rangle^{\otimes 2} .
 \end{eqnarray}
 Under the same assumptions as before, we want to discriminate the density matrices
 \begin{equation}
 \rho_{12} = \frac{1}{3} \sum_{j=1}^{3} |\mu_{j}\rangle\langle\mu_{j}| ,
 \end{equation}
 and
 \begin{equation}
 \rho_{34} = \frac{1}{3} \sum_{j=4}^{6} |\mu_{j}\rangle\langle\mu_{j}| .
 \end{equation}
 In order to apply Helstrom's method, we have to diagonalize $\rho_{12} - \rho_{34}$, and this is a $6$ dimensional problem.  One way to proceed is to express the eigenstates as linear combinations of the linearly independent vectors $\{ |\mu_{j}\rangle\, |\, j=1,2,\ldots 6\}$, $|\zeta\rangle = \sum_{j=1}^{6}d_{j}|\mu_{j}\rangle$.  The equation $( \rho_{12} - \rho_{34} )|\zeta\rangle = \lambda |\zeta\rangle$ can be expressed as $M|\zeta\rangle = \lambda |\zeta\rangle$, where the matrix $M$, which is expressed in a non-orthonormal basis, is
 \begin{equation}
 M= \left( \begin{array}{cccccc} 1 & c^{2} & c & c^{2} & c^{2} & c^{3} \\ c^{2} & 1 & c & c^{2} & c^{2} & c^{3} \\ c & c & 1 & c^{3} & c^{3} & c^{4} \\ -c^{2} & -c^{2} & -c^{3} & -1 & -c^{2} & -c \\ -c^{2} & -c^{2} & -c^{3} & -c^{2} & -1 & -c \\ -c^{3} & -c^{3} & -c^{4} & -c & -c & -1 \end{array} \right) ,
 \end{equation}
where $c=\cos\theta$.  This matrix can be diagonalized using Mathematica, and the results inserted into the Helstrom formula for the success probability.  The success probability is plotted for both the entangled state and the product state in Fig. 2.  As can be seen the entangled state performs better for values of $\theta$ less than approximately $0.7$, and this corresponds to the weak signal regime, which is the one of greatest interest.

\begin{figure}
\centering
\includegraphics[width=0.5\textwidth]{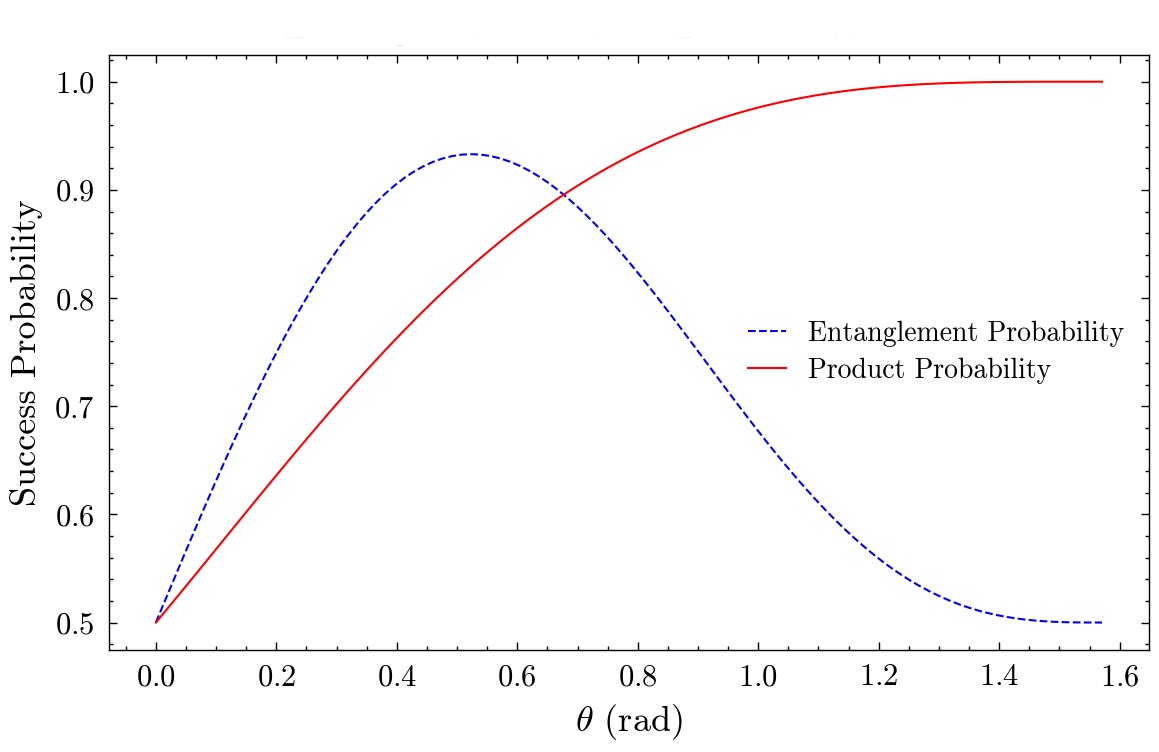}
\caption{Comparison of probability of successfully discriminating between excitations in two groups of detectors with an entangled initial state and with a product initial state.  For small $\theta$, corresponding to weak signals, the entangled state does better.}
\end{figure}

 Since the entangled state was better at discriminating which set of detectors fired for small theta, let us now expand on it further.  As previously noted, we are most interested in the case of weak signals, which correspond to small $\theta$.  We will add a third state, the one that corresponds to no detector firing.  That state is just $\rho_{0}= |\psi_{in}\rangle\langle\psi_{in}|$, or, in matrix form
\begin{equation}
\rho_{0}= \frac{1}{2} \left( \begin{array}{cc} 1 & 1 \\ 1 & 1 \end{array} \right) .
\end{equation}
The idea is to discriminate between $\rho_{0}$, $\rho_{12}$, and $\rho_{34}$, and we will assume each occurs with a probability of $1/3$.  In order to do this, we will use the pretty good measurement.  

The pretty-good measurement allows us to find a good approximation to the minimum-error POVM for discriminating general sets of states \cite{hausladen}.  If we wish to discriminate the states $\rho_{j}$, $j=1,2,\ldots N$, and $\rho_{j}$ occurs with probability $p_{j}$, then the pretty-good measurement gives the POVM elements
 \begin{equation}
 \Pi_{j}=p_{j} \rho^{-1/2} \rho_{j} \rho^{-1/2} ,
 \end{equation}
where $\rho = \sum_{j=1}^{N} p_{j} \rho_{j}$, and the probability of successfully discriminating the states is
 \begin{equation}
 P_{s}=\sum_{j=1}^{N} p_{j} {\rm Tr}(\Pi_{j}\rho_{j}) .
\end{equation}
Applying this to the current case, we have that the POVM elements are 
\begin{eqnarray}
\Pi_{0} & = & \frac{1}{3} \rho^{-1/2} \rho_{0} \rho^{-1/2} \nonumber \\
\Pi_{12} & = & \frac{1}{3} \rho^{-1/2} \rho_{12} \rho^{-1/2}  \nonumber \\
\Pi_{34} & = & \frac{1}{3} \rho^{-1/2} \rho_{34} \rho^{-1/2} ,
\end{eqnarray} 
where $\rho = (1/3)(\rho_{0} + \rho_{12} + \rho_{34})$, or
\begin{equation}
\rho = \left( \begin{array}{cc} \frac{1}{2} & a \\ a & \frac{1}{2} \end{array} \right) ,
\end{equation}
where $a = \frac{1}{6} + \frac{2}{9} \cos (2\theta ) + \frac{1}{9} \cos (4\theta )$. The eigenvalues of $\rho$ are $(1/2)\pm a$ and the eigenvectors are
\begin{equation}
\frac{1}{\sqrt{2}} \left( \begin{array}{c} 1 \\ 1 \end{array} \right) \hspace{1cm} \frac{1}{\sqrt{2}} \left( \begin{array}{c} 1 \\ -1 \end{array} \right). 
\end{equation}
We can then write
\begin{eqnarray}
\rho^{-1/2} & = & \frac{1}{2[(1/2)+ a]^{1/2}} \left( \begin{array}{cc} 1 & 1 \\ 1 & 1 \end{array} \right) \nonumber \\
&& + \frac{1}{2[(1/2)- a]^{1/2}} \left( \begin{array}{cc} 1 & -1 \\ -1 & 1 \end{array} \right) .
\end{eqnarray}
The POVM elements are
\begin{equation}
\Pi_{0} = \frac{1}{6} \frac{1}{(1/2)+a} \left( \begin{array}{cc} 1 & 1 \\ 1 & 1 \end{array} \right) ,
\end{equation}
\begin{eqnarray}
\Pi_{12} & = & \frac{1}{12} \left[ \frac{1}{(1/2)+a} \left( \begin{array}{cc} 2(1+3a) & -1 \\-1 & 2(1+3a)\end{array} \right)  \right. \nonumber \\
&& \left. + \frac{2i}{3}  \frac{\left( 2\sin (2\theta ) + \sin (4\theta ) \right)}{ [(1/4) - a^{2}]^{1/2}} \left( \begin{array}{cc} 0 & 1 \\ -1 & 0 \end{array} \right) \right] ,
\end{eqnarray} 
and $\Pi_{34}= \Pi_{12}^{\ast}$.  

The success probability for this measurement is given by
\begin{equation}
P_{s} = \frac{1}{3} \left[ {\rm Tr}(\Pi_{0}\rho_{0}) + {\rm Tr}(\Pi_{12}\rho_{12}) + {\rm Tr}(\Pi_{34}\rho_{34}) \right] .
\end{equation}
We find that ${\rm Tr}(\Pi_{12}\rho_{12}) = {\rm Tr}(\Pi_{34}\rho_{34})$ and
\begin{eqnarray}
{\rm Tr}(\Pi_{0}\rho_{0}) & = & \frac{1}{(3/2)+ 3a}  \nonumber \\
{\rm Tr}(\Pi_{12}\rho_{12}) & = & \frac{3a+(5/2)}{6+12a} \nonumber \\
&&+ \frac{1}{6[(1/4)-a^{2}]^{1/2}} \left[\frac{2}{3} \sin (2\theta ) + \frac{1}{3}\sin (4\theta )\right]^{2}  .\nonumber \\
\end{eqnarray}
This is plotted in figure 3.  Note that the success probability is lower that it was in the case when the no-firing state was not included.  This is to be expected, since increasing the set of states to be discriminated increases the difficulty of the task.

\begin{figure}
\centering
\includegraphics[width=0.5\textwidth]{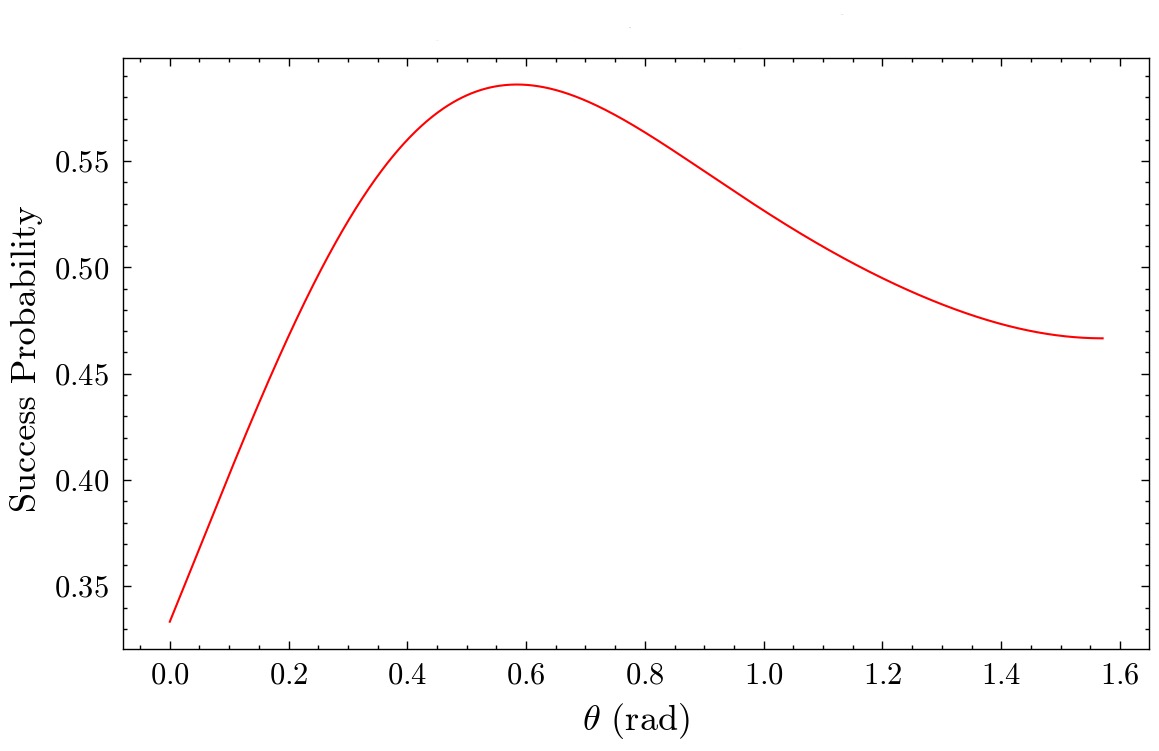}
\caption{Success probability of discriminating the detector states in which the case that neither group of detectors fires is included.}
\end{figure}

\section{Unambiguous discrimination and clusters of detectors}
We already looked at the unambiguous discrimination of states of two detectors, and now we want to extend those results to two clusters of two detectors.  We are able to build on our previous results to do this, and the form of the solutions suggests how they can be extended to more than two clusters.  The object of this section is to compare the results of a collective measurement to those of measuring each detector individually.

We will start the system in the state $|\psi_{in}\rangle = |\phi_{00}\rangle_{12}\otimes |\phi_{00}\rangle_{34}$.  Let 
\begin{equation}
R=I-|\phi_{00}\rangle\langle\phi_{00}| .
\end{equation}
Note that $R$ is a projection and that $R|\phi_{00}\rangle = 0$.  We now define POVM elements
\begin{eqnarray}
\Pi_{0} & = & c_{0} |\phi_{00}^{\perp}\rangle_{12}\langle\phi_{00}^{\perp}| \otimes |\phi_{00}^{\perp}\rangle_{34}\langle\phi_{00}^{\perp}| \nonumber \\
\Pi_{12} & = & c_{1} R_{12} \otimes |\phi_{00}^{\perp}\rangle_{34}\langle\phi_{00}^{\perp}| \nonumber \\
\Pi_{34} & = & c_{1} |\phi_{00}^{\perp}\rangle_{12}\langle\phi_{00}^{\perp}| \otimes R_{34} \nonumber \\
\Pi_{both} & = & c_{2} R_{12} \otimes R_{34} .
\end{eqnarray}
The element $\Pi_{0}$ corresponds to no detectors being excited, $\Pi_{12}$ corresponds to only detectors in cluster $12$ being excited, $\Pi_{34}$ corresponds to only detectors in cluster $34$ being excited, and $\Pi_{both}$ corresponds to detectors in both clusters being excited.  The constants $c_{0}$, $c_{1}$, and $c_{2}$ are determined by the condition that the failure operator,
\begin{equation}
\Pi_{f} = I_{12}\otimes I_{34} - \Pi_{0} - \Pi_{12} - \Pi_{34} - \Pi_{both} ,
\end{equation}
be a positive operator.

 Now let us split the individual Hilbert spaces into two orthogonal subspaces, $S={\rm span}\, \{ |\phi_{00}\rangle , |\phi_{00}^{\perp}\rangle \}$, and $S^{\perp}$, which consists of all vectors orthogonal to $S$.  Now consider a vector in $S_{12}^{\perp}\otimes S_{34}^{\perp}$, which is a four-dimensional space.  When acting on this vector, only the first and last term of $\Pi_{f}$ give a nonzero contribution, and that contribution is $1-c_{2}$ times the vector.  Therefore, $\Pi_{f}$ has $1-c_{2}$ as an eigenvalue and this eigenvalue is four-fold degenerate.  The condition that this eigenvalue be greater than or equal to zero is just $1\geq c_{2}$.  Now consider a vector in $S_{12}\otimes S_{34}^{\perp}$, also a four-dimensional space.  Only the first, fourth, and fifth terms of $\Pi_{f}$ give nonzero contributions, and note that $\Pi_{f}$ acting on a vector in $S_{12}\otimes S_{34}^{\perp}$ maps the vector into $S_{12}\otimes S_{34}^{\perp}$.  That means we can find 4 more eigenvalues of $\Pi_{f}$ by diagonalizing $\Pi_{f}$ in the space $S_{12}\otimes S_{34}^{\perp}$.  Doing so is relatively straightforward.  Let $|v\rangle \in S^{\perp}$.  Then
\begin{eqnarray}
\Pi_{f} |\phi_{00}\rangle_{12}|v\rangle_{34} & = & |\phi_{00}\rangle_{12}|v\rangle_{34} - c_{1} \langle \phi_{00}^{\perp}|\phi_{00}\rangle\,  |\phi_{00}^{\perp}\rangle_{12}|v\rangle_{34} \nonumber \\
\Pi_{f} |\phi_{00}^{\perp}\rangle_{12}|v\rangle_{34} & = & |\phi_{00}^{\perp}\rangle_{12}|v\rangle_{34} -c_{1} |\phi_{00}^{\perp}\rangle_{12}|v\rangle_{34}  -c_{2} (|\phi_{00}^{\perp}\rangle \nonumber \\
&& - \langle\phi_{00}|\phi_{00}^{\perp}\rangle |\phi_{00}\rangle_{12}) |v\rangle_{34}  , 
\end{eqnarray}
and we see that since the vector $|v\rangle_{34}$ is not affected by this transformation, all we really have to do is diagonalize the part of it in $S$.  If we define our eigenstate to be $|\zeta\rangle = d_{0}|\phi_{00}\rangle + d_{1}|\phi_{00}^{\perp}\rangle$, we want it to satisfy $\Pi_{f} |\zeta\rangle_{12}|v\rangle_{34} = \lambda   |\zeta\rangle_{12}|v\rangle_{34}$, and the above equation then implies
\begin{eqnarray}
d_{0} + d_{1}c_{2} z^{\ast} & = & \lambda d_{0} \nonumber \\
-d_{0}c_{1}z + d_{1}(1-c_{1}-c_{2}) & = & \lambda d_{1} ,
\end{eqnarray}
where $z=\langle\phi_{00}^{\perp}|\phi_{00}\rangle = -\sin^{2}\theta$.  These equations can be put in matrix form yielding the characteristic equation for the eigenvalues
\begin{equation}
\lambda^{2} -\lambda (2-c_{1}-c_{2}) + 1-c_{1}-c_{2}+c_{1}c_{2}|z|^{2} = 0 .
\end{equation}
For both eigenvalues to be positive, we need $2-c_{1}-c_{2} \geq 0$ and $1-c_{1}-c_{2}+c_{1}c_{2}|z|^{2} \geq 0$.  Note that if the second condition is satisfied the first is too, so we really only have the second condition here.

Similarly, we diagonalize the part of $\Pi_{f}$ in $S_{12}^{\perp}\otimes S_{34}$, and this yields the same condition on $c_{1}$, $c_{2}$, and $c_{3}$ as above.

Now let's diagonalize  $\Pi_{f}$ in  $S_{12}\otimes S_{34}$, which is the span of $ |\phi_{00}\rangle_{12} |\phi_{00}\rangle_{34}$, $|\phi_{00}\rangle_{12} |\phi_{00}^{\perp}\rangle_{34}$, $|\phi_{00}^{\perp}\rangle_{12} |\phi_{00}\rangle_{34}$, and $|\phi_{00}^{\perp}\rangle_{12} |\phi_{00}^{\perp}\rangle_{34}$.  We find that
\begin{eqnarray}
\Pi_{f}|\phi_{00}\rangle_{12} |\phi_{00}\rangle_{34} & = & |\phi_{00}\rangle_{12} |\phi_{00}\rangle_{34} -c_{0} z^{2} |\phi_{00}^{\perp}\rangle_{12} |\phi_{00}^{\perp}\rangle_{34} \nonumber \\
\Pi_{f} |\phi_{00}\rangle_{12} |\phi_{00}^{\perp}\rangle_{34} & = & |\phi_{00}\rangle_{12} |\phi_{00}^{\perp}\rangle_{34} - z(c_{0}+c_{1})|\phi_{00}^{\perp}\rangle_{12} |\phi_{00}^{\perp}\rangle_{34} \nonumber \\
&&+ c_{1} |z|^{2}   |\phi_{00}^{\perp}\rangle_{12} |\phi_{00}\rangle_{34} \nonumber \\
\Pi_{f}  |\phi_{00}^{\perp}\rangle_{12} |\phi_{00}\rangle_{34} & = &  |\phi_{00}^{\perp}\rangle_{12} |\phi_{00}\rangle_{34} -z(c_{0}+c_{1})  |\phi_{00}^{\perp}\rangle_{12} |\phi_{00}^{\perp}\rangle_{34} \nonumber \\
&&+ c_{1}|z|^{2} |\phi_{00}\rangle_{12} |\phi_{00}^{\perp}\rangle_{34} , \nonumber \\
\end{eqnarray}
and
\begin{eqnarray}
\Pi_{f} |\phi_{00}^{\perp}\rangle_{12} |\phi_{00}^{\perp}\rangle_{34} & = & (1-c_{0}-2c_{1}-c_{2})|\phi_{00}^{\perp}\rangle_{12} |\phi_{00}^{\perp}\rangle_{34} \nonumber \\
&&+ z^{\ast}(c_{1}+c_{2})(|\phi_{00}\rangle_{12} |\phi_{00}^{\perp}\rangle_{34} \nonumber \\
&&+ |\phi_{00}^{\perp}\rangle_{12} |\phi_{00}\rangle_{34} ) - c_{2}(z^{\ast})^2 |\phi_{00}\rangle_{12} |\phi_{00}\rangle_{34} . \nonumber \\
\end{eqnarray}
Expressing the eigenvector of $\Pi_{f}$ confined to $S_{12}\otimes S_{34}$ as
\begin{eqnarray}
|\zeta\rangle & = & d_{0} |\phi_{00}\rangle_{12} |\phi_{00}\rangle_{34} + d_{1} |\phi_{00}\rangle_{12} |\phi_{00}^{\perp}\rangle_{34} \nonumber \\
&& + d_{2} |\phi_{00}^{\perp}\rangle_{12} |\phi_{00}\rangle_{34} + d_{3} |\phi_{00}^{\perp}\rangle_{12} |\phi_{00}^{\perp}\rangle_{34} ,
\end{eqnarray} 
we find the eigenvalue equation $\Pi_{f}|\zeta\rangle = \lambda |\zeta\rangle$ can be expressed as
\begin{equation}
L \left(\begin{array}{c} d_{0} \\d_{1} \\ d_{2} \\ d_{3} \end{array} \right) 
= \lambda \left(\begin{array}{c} d_{0} \\d_{1} \\ d_{2} \\ d_{3} \end{array}\right)
\end{equation}
where the matrix $L$ is given by
\begin{equation}
L= \left( \begin{array}{cccc} 1 & 0 & 0 & -(z^{\ast})^{2} \\ 0 & 1 & c_{1}|z|^{2} & z^{\ast}(c_{1}+c_{2}) \\ 0 & c_{1}|z|^{2} & 1 & z^{\ast}(c_{1}+c_{2}) \\ -c_{0}z^{2} & -z(c_{0}+c_{1}) & -z(c_{0}+c_{1}) & 1-c_{0}-2c_{1}-c_{2} \end{array} \right)  .
\end{equation}
Defining $x=1-\lambda$, the characteristic equation of this matrix is
\begin{eqnarray} 
0 & = & x^{4} - (c_{0}+2c_{1}+c_{2}) x^{3} + |z|^{2}[ 2(c_{0}+c_{1})(c_{1}+c_{2}) \nonumber \\
&&- |z|^{2}(c_{2}c_{0}+c_{1}^{2})] x^{2} + |z|^{4}c_{1}(c_{1}^{2}-c_{0}c_{2}) x \nonumber \\
&&+ |z|^{8}c_{0}c_{2}c_{1}^{2} ,
\end{eqnarray}
and the condition that the eigenvalues be nonnegative is equivalent to the condition that the roots of this equation satisfy $x\leq 1$.  To solve this equation, we will adopt a perturbative approach.  For $\theta$ small, $|z|$ is also small, so keeping only terms up to first order in $|z|^{2}$, we find two roots of $x=0$, and, setting $b=c_{0}+2c_{1}+c_{2}$, the two additional roots
\begin{equation}
x=\frac{1}{2}\{ b \pm [b^{2} -8|z|^{2}(c_{0}+c_{1})(c_{1}+c_{2})]^{1/2}\} < b .
\end{equation}
Therefore, if $b\leq 1$, the roots will satisfy $x\leq 1$.  Taking into account all the conditions on $c_{0}$, $c_{1}$, and $c_{2}$, the choice $c_{0}=c_{1}=c_{2}=1/4$ will guarantee that the operator $\Pi_{f}$ is nonnegative.  This solution is not guaranteed to be optimal, but it is sufficient to demonstrate that the collective measurement outperforms individual measurements.

Now that we have the POVM we can calculate its success probability, assuming each of the states $|\phi_{jk}\rangle_{12} \otimes |\phi_{lm}\rangle_{34}$, $j,k,l,m = 0,1$ is equally likely.  We find that
\begin{eqnarray}
P_{s} & = & \frac{1}{16} \sin^{4}\theta \left[ \left(1 + \frac{1}{2}\sin^{2}\theta \right)^{2} \right. \nonumber \\
&& \left. + (1+\cos^{2}\theta)\left( \frac{3}{2} + \frac{1}{4}\sin^{2}\theta \right) \right] .
\end{eqnarray}
This can be compared to the case that each detector is measured individually.  For that measurement to succeed, all four of the individual measurements must succeed.  This will happen with a probability of (see Eq.\ (\ref{unam-n}) )
\begin{equation}
P_{s-ind} = \left( \frac{\sin^{2}\theta}{1+\cos\theta} \right)^{4} .
\end{equation}
These are plotted in Fig.\ 4, and it can be seen that while both curves are close to one for small $\theta$, the failure probability of the collective POVM measurement is lower than that provided by measuring the detectors individually.  While the POVM measurement is not optimal, the fact that it does better than individual measurements, does show that there is an advantage to be gained by collective measurements.

\begin{figure}
\label{figure4}
\centering
\includegraphics[width=0.48\textwidth]{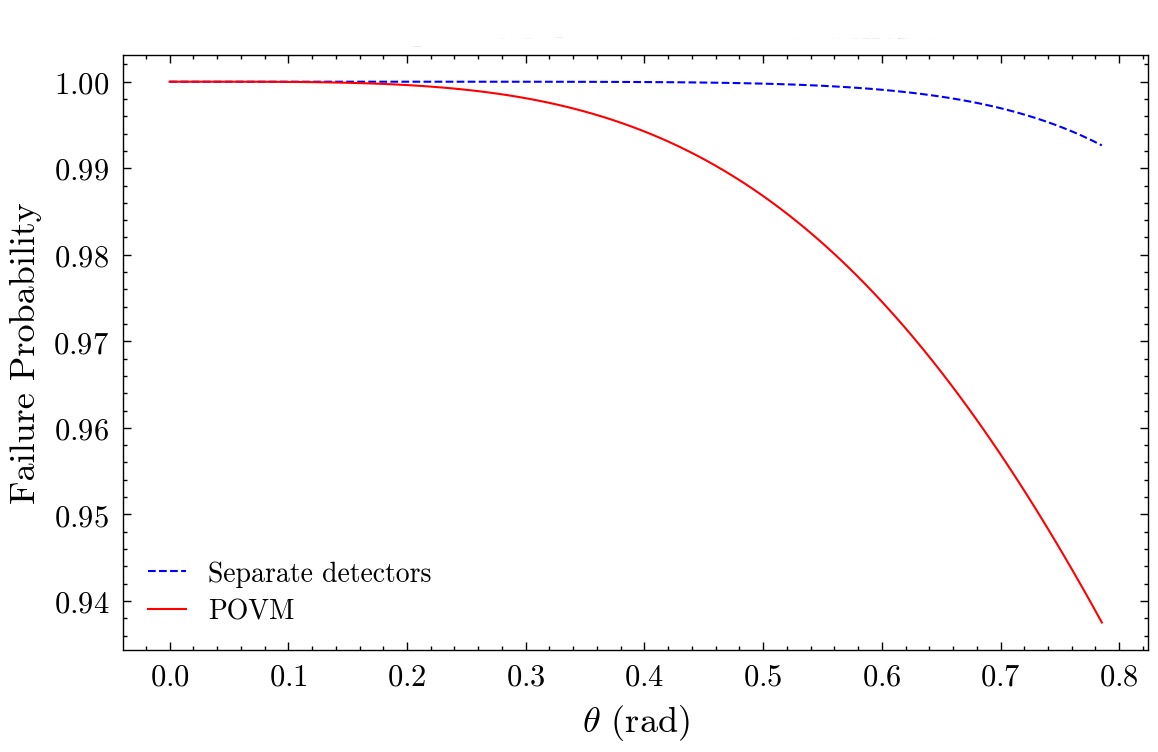}
\caption{Failure probability, $P_{f}$ as a function of $\theta$ for the unambiguous discrimination of two groups of detectors.  The solid curve is for the POVM and the dashed curve is for measuring each detector individually.  As before the failure probability remains close to one for a significant range of $\theta$, but the POVM performs better than individual measurements.}
\end{figure}

\section{Conclusion}
We have examined a number of different scenarios involving detector networks.  We began with only two detectors, and our conclusion was that the best initial state of the detectors when we wanted to find which detectors fired was a product state.  We then moved on to arranging the detectors into groups, and here entangled initial states proved useful.  We considered two groups each consisting of two detectors, and we only wished to know in which group detectors fired, and not which individual detectors did so.  The case in which only one detector fires, can be reduced to that of only two detectors, and an entangled initial state is best.  In the case in which one or two detectors can fire in each group, we studied both entangled and product initial states, and for small interaction strengths, the entangled state was better.  This was extended to include the case that detectors in neither group fire.  Most of these studies made use of minimum error discrimination.  Finally, we looked at the case of unambiguous discrimination with an initial product state, and compared a collective POVM measurement to measuring each detector individually.  The POVM measurement was found to perform better.  Our overall conclusion is that when considering groups of detectors and only requiring information about in which group a detection occurred, entangled initial states and collective measurements can provide an advantage.

\acknowledgements
This research was supported by the NSF under grant FET-2106447.

\end{document}